\documentclass[
twocolumn,
groupedaddress,
superscriptaddress,
amsmath,
amssymb,
amsfonts, 
amstext,
]{revtex4-2}

\usepackage{dcolumn}
\usepackage{bm}
\usepackage{soul}
\usepackage[flushleft]{threeparttable}
\usepackage{url}
\usepackage{mathrsfs}
\usepackage{braket}
\usepackage{multirow}
\usepackage{array}
\usepackage{booktabs}
\usepackage{float}
\usepackage{lineno}
\usepackage{hyperref}
\usepackage{mathtools}
\usepackage{xcolor}
\usepackage{MnSymbol}
\usepackage{enumitem}
\usepackage{tikz}
\newcommand*{\circled}[1]{\lower.7ex\hbox{\tikz\draw (0pt, 0pt)%
    circle (.42em) node {\makebox[1em][c]{\small #1}};}} 

\hypersetup{colorlinks=true, urlcolor=blue, citecolor=blue, linkcolor=blue, pdfborder={0 0 0}}

\begin{document}



\title{Accurate polymorphous description of the paramagnetic phases in MnBi$_{2}$Te$_{4}$}

\author{Yufei Zhao}
\affiliation{%
Department of Physics, Southern University of Science and Technology, Shenzhen 518055, China
}%
\author{Qiushi Yao}
\affiliation{%
Department of Physics, Southern University of Science and Technology, Shenzhen 518055, China
}%
\author{Pengfei Liu}
\affiliation{%
Department of Physics, Southern University of Science and Technology, Shenzhen 518055, China
}%
\author{Qihang Liu}
\email{liuqh@sustech.edu.cn}
\affiliation{%
Department of Physics, Southern University of Science and Technology, Shenzhen 518055, China
}%
\affiliation{%
Guangdong Provincial Key Laboratory for Computational Science and Material Design, Southern University of Science and Technology, Shenzhen 518055, China
}%
\affiliation{%
Shenzhen Key Laboratory of Advanced Quantum Functional Materials and Devices, Southern University of Science and Technology, Shenzhen 518055, China
}%

\begin{abstract}
Temperature-driven phase transition is a long-standing frontier in material science, among which the most common phenomenon is the transition from a low-temperature magnetic-ordered phase to a high-temperature paramagnetic phase. A paramount question is if such a paramagnetic phase of the “correlated solids” can be well described by single-particle band theory to facilitate the experimental observations. In this work, we investigate the electronic properties of the paramagnetic phase by the static density functional theory via two different approaches, namely monomorphous description and polymorphous description. In the conventional monomorphous description, the local spin moments are naively forced to be zero. By contrast, the polymorphous description based on a large enough supercell with disordered distributed local moments is able to count in the effects of distinct local environments, providing a more reliable paramagnetic electronic structure to simulate realistic materials. From a comparison of total energies, symmetries, and band structures, we demonstrate the necessity for a proper treatment of paramagnetic phases, taking MnBi$_{2}$Te$_{4}$ as an example. Our work provides a theoretical perspective on the evolution of electronic structures through magnetic order-disorder phase transitions in emergent topological magnets.
\end{abstract}

\maketitle


\section{Introduction}
Time-reversal ($\mathcal{T}$) symmetry is known to play a primary role in the characteristics of topological phases of matter \cite{RevModPhys.82.3045,RevModPhys.83.1057}. Its presence or absence is a bedrock for the classification and achievement of emergent quantum phenomena, such as quantum spin Hall (QSH) effect in a $\mathcal{T}$-preserved topological insulator \cite{2006quantum} and quantum anomalous Hall (QAH) effect in a $\mathcal{T}$-broken Chern insulator \cite{haldane1988model}. In the past few years, experiment and theory have established elegant consistency in $\mathcal{T}$-broken systems and unearthed a series of ordered magnetic topological insulators and Weyl semimetals with the long-range magnetic order \cite{chang2013experimental, li2019dirac,YLChen, PhysRevX.11.011039, doi:10.1126/science.aav2873, liu2020robust, PhysRevLett.122.107202, PhysRevB.103.L121112, tan2022mnbi2te4, xu2022visualization}. However, a long-standing problem in density-functional theory (DFT) is how to deal with paramagnetic (PM) phases with disordered distributed local moments, where $\mathcal{T}$ is globally preserved due to macroscopically zero net spin moments but locally broken around the magnetic ions \cite{zunger2022bridging}.

Before reviewing this problem, it should be noted that the evolution of the electronic band structure with temperature provides direct evidence of magnetism-induced topological phase transitions. For instance, the surface gap of a magnetic topological insulator is the prerequisite for realizing the QAH effect and axion response \cite{10.1063/5.0059447}. Such a gap originates from magnetism (rather than other effects such as Coulomb scattering \cite{dephasing}) only if it disappears with temperature $T$ past the critical temperatures $T_{C}$ or $T_{N}$ \cite{PhysRevX.11.011039}. Another expected possible scenario is in magnetic Weyl semimetal, whether two Weyl points merge to a Dirac point or are gapped to an insulator \cite{RevModPhys.90.015001}. Observing these electronic fine changes cannot be done by solely tracing any ``kinks'' in other order parameters, such as heat capacity $C(T)$ or susceptibility $\chi(T)$.

Conventionally, the PM phase from first-principles calculations is done by simply forcing the local moment of each magnetic atom ($e.g.$, transition-metal atom) to be homogenously zero for a macroscopic statistical behavior, namely a nonmagnetic (NM) monomorphous description \cite{RN39,RN40,RN42,RN99,li2019dirac}. In other words, the magnetic cations are symmetry-equivalent and share an identical local environment. Although such an approximation yields reasonable results in some scenarios, it usually leads to severe discrepancies with the physical reality and experiments. One may note that a broad range of Mott insulators and perovskites with 3$d$ electrons have been reported to maintain their bandgap in both low-$T$ and high-$T$ phases, but however, these compounds exhibit false-positive metallic states under the NM monomorphous calculations \cite{Trimarchi,RN83,PhysRevB.102.045112,MALYI202035,PhysRevB.102.214417}. As a result, the mean-field DFT was once claimed to be insufficient for describing the correlated PM phases.

Indeed, when the local moments of the magnetic atoms are randomly aligned at $T > T_{C}$, the resultant PM phase has an ensemble-averaged property $\braket{P(S^{i})}$ ($e.g.$, band structure) from numerous random magnetic configurations $S^{i}$. By contrast, the monomorphous approach, $i.e.$ the NM model within a single minimal primitive cell, yields a property $P(\braket{S^{i}})$ of artificially averaged magnetic structure $\braket{S^{i}}$ at each atomic site \cite{MALYI202035,zunger2022bridging,PhysRevMaterials.5.024207}. Consequently, spurious $\mathcal{T}$ symmetry (both globally and locally) and crystalline symmetries lead to artificial results such as partial filling of bands and thus problematic predictions of the wavefunctions as well as the topological properties.

This article attempts to answer a simple yet undeniable question: How to identify the magnetic order$-$disorder phase transition directly from DFT-calculated band structures? We address it by applying the state-of-the-art polymorphous description to offer a prediction for the bulk and thin films. The rest of the paper is organized as follows. Section \ref{method} provides the methodology adopted and computational details. In Section \ref{results}, the intrinsic magnetic topological insulator MnBi$_{2}$Te$_{4}$ is taken as a typical example. Through the polymorphous calculations, we find exclusive advantages of this method in describing the band structure and total energy, which is well consistent with the previous experiments. Our results establish an important routine for capturing the transition signals, in order to compare with angle-resolved photoemission spectroscopy (ARPES) and scanning tunneling microscope (STM) measurements.

\section{\label{method}Methods}
Our DFT calculations are carried out by using the projector augmented-wave (PAW) method, implemented in Vienna \textit{ab-initio} Simulation Package (\textsc{vasp}) \cite{RN17,RN15,PhysRevB.50.17953}. The Perdew-Burke-Ernzerhof (PBE) type exchange-correlation functional in the generalized gradient approximation (GGA) \cite{RN16} is adapted to take into account exchange and correlation contributions to the Hamiltonian of the electron-electron system. Since the on-site Coulomb interactions among electrons on Mn-3$d$ are strong, we have taken $U$ = 5.0 eV as a parameter in the GGA+$U$ calculations \cite{PhysRevB.57.1505}. We apply the DFT-D3 approach \cite{rn19} to describe the van der Waals interactions. To account for the effect of fluctuated magnetic moments on the total energy, the coordinates of the atoms are optimized (only for the bulk). Energy cost as a function of magnetic moment is performed via the constrained density functional theory (CDFT) \cite{kaduk2012constrained}. In order to obtain an intuitive electronic spectrum rather than confused heavily folded bands, we apply a rigorous band unfolding \cite{RN21,RN32} to obtain an effective band structure (EBS) in a primitive Brillouin zone,
\begin{equation}
    P_{\Vec{K}m} (\Vec{k}_{i}) = \sum_{n} \vert \braket{\Vec{K}m | \Vec{k}_{i}n} \vert ^{2},
\end{equation}
where $\ket{\Vec{K}m}$ and $\ket{\Vec{k}n}$ are the eigenvectors of the supercell and primitive cell. $\ket{\Vec{K}m}$ can be expressed as a linear combination of primitive cell eigenvectors $\ket{\Vec{k_i}n}$. The spectral weight is given by
\begin{equation}
    A(\Vec{k_i}n, E) = \sum_{m} P_{\Vec{K}m} (\Vec{k}_{i})\delta(E_{m} - E).
\end{equation}
This process is implemented in the open-source code \textsc{bandup} \cite{BANDUP1,BANDUP2}. 

We next introduce how to polymorphously simulate the disordered PM phase by applying the “special quasirandom structures” (SQSs) method \cite{PhysRevLett.65.353}. Instead of averaging the band structures of many PM snapshot configurations $\{ S^{i} \}$, which belongs to another form of polymorphous description but is clumsy and inefficient, we construct a single but large enough supercell that takes into account the individual local moments as well as the local disordered effects. It provides more reliable results than the ensemble average along many small random supercells. From statistical mechanics, the guidelines for picking the desired supercell are as follows:

($i$) \textit{Multisite correlation function.}  We first define a configuration $S^{i}$ represented by a vector of occupation variables $S^{i}_{j}$, indicating which type of atom sits on lattice site $j$. In a multi-component system ($e.g.$ alloy), $M_{j}$ distinct chemical species can be assigned to occupy site $j$ (denoted as $S^{i}_{j} = 0, \cdots, M_{j}-1$ due to its chemical composition). In most cases, the PM phase can be approximately treated as a binary alloy with two species: spin-up and spin-down. The reasons will be clarified later. We also need to define the ``cluster'' $\alpha^{\kappa, \xi}$ to characterize the $\kappa$-order arrangement of $\xi$th-nearest neighbor atoms, $e.g.$ points $\alpha^{1}$, lines $\alpha^{2}$, triangles $\alpha^{3}$. For simplicity, $\xi$ is omitted hereafter.

For a given configuration $S$, the multisite correlation function of a specific cluster $\alpha^{\kappa}$ is given by
\begin{equation}
    \Gamma_{\alpha^{\kappa}}(S) = \prod_{i}\gamma_{\alpha^{\kappa}}(S_{i}).
\end{equation}
In a binary alloy, a common choice of $\gamma$ is $\gamma(0) = +1$ and $\gamma(1) = -1$. If we only consider $\kappa=2$, it is the Ising model. 

The correlation of the whole system is expressed as 
\begin{equation}
    \rho(\sigma) = \sum_{\alpha} D_{\alpha} m_{\alpha} \braket{\Gamma_{\alpha^{\prime}}(\bm{\sigma})}_{\alpha^{\kappa}},
\end{equation}
where $\braket{\Gamma_{\alpha^{\prime}}(\bm{\sigma})}_{\alpha}$ sums over all the clusters $\alpha^{\prime}$ which are symmetry equivalent to $\alpha^{\kappa}$. $m_{\alpha}$ is the number of clusters $\alpha^{\prime}$. $D_{\alpha}$ is cluster expansion coefficients or cluster interaction potential. In the fully disordered PM phase with complete stochasticity, the correlation function is zero $\rho_{\alpha}(S^{*}) = 0$. To search for a supercell with a minimum correlation approaching $\rho_{\alpha}(S^{*})$, we apply a Monte-Carlo method \textsc{mcsqs} implemented in Alloy Theoretic Automated Toolkit (\textsc{atat}) \cite{SQS1, SQS2}. In this paper, the cutoff radii $\xi$ of pair correlation ($\kappa=2$) and triplet correlation ($\kappa=3$) clusters are set between the second and third nearest neighbors.

($ii$) \textit{Spin orientation.} In the language of the classical Heisenberg model, at low temperatures, the magnetic moments interact with each other through (direct or super) exchange interaction and tend to align parallel (ferromagnetic, FM) or antiparallel (antiferromagnetic, AFM) to reach the lower-energy ground state, expressed as \cite{RN99}
\begin{equation}
    \hat{H} = -\sum_{i, j} J_{ij} \sigma_{i} \sigma_{j} = -\sum_{\alpha} J_{\alpha} n_{\alpha} \langle \Phi_{\alpha} \rangle
\end{equation}
where $\langle \Phi_{\alpha} \rangle = \frac{1}{N} \sum_{i,j\in\alpha} \sigma_{i} \sigma_{j}$ is the average spin-spin correlation of a given type of cluster $\alpha$. $J_{\alpha}$ and $n_{\alpha}$ denote the coupling constant and the number of atoms. Consider a one-dimensional Ising model, the energies of the FM and AFM phases satisfy ${E_{\rm FM}} = -{E_{\rm AFM}} = -\sum_{\alpha}J_{\alpha}$.  Because the averaged correlation function should fulfill $\langle \Phi_{\alpha} \rangle =0, \forall \alpha$ in the disordered PM phase, ${E_{\rm PM}} = 0$ is consistent with the requirement ($i$).

\begin{figure*}[t]
\includegraphics[scale=0.6]{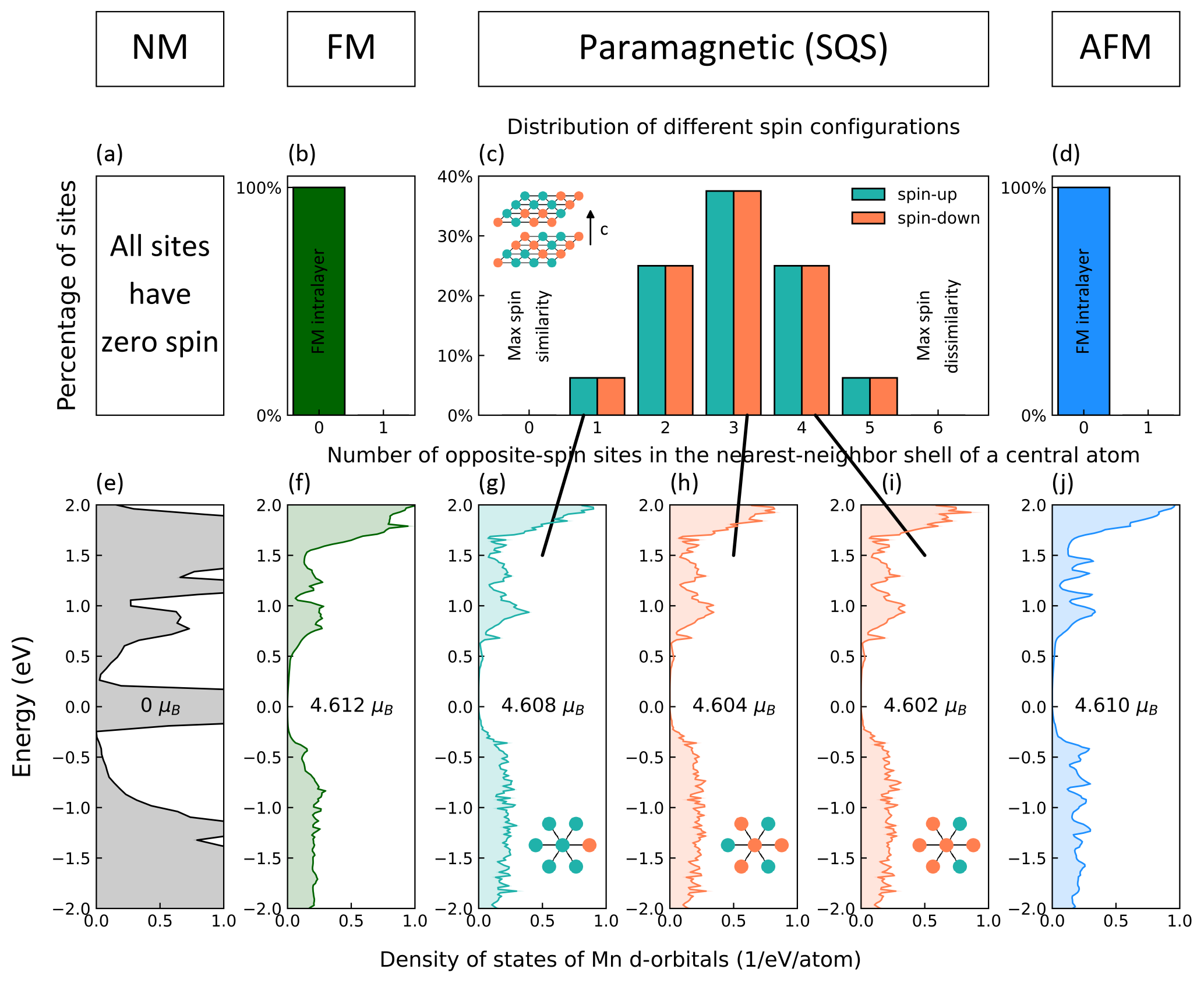}
\caption{\label{fig1}The spin configuration and corresponding projected local density of states (PDOS) in MnBi$_2$Te$_4$. (a)-(d) Upper panel: The weight of local spin configuration $F(n^{\uparrow\downarrow})$ in different spin configurations. (a) All the metal sites are identical for zero spins. (b) and (d) The spins of nearest neighbor cations in a layer, keep the same direction in the FM and $A$-type AFM phases ($n^{\uparrow\downarrow}= 0$ everywhere). (c) Distribution of the weights of local spin configurations in the PM supercell modeled by the 224-atom SQS shown in the inset (only the Manganese sites are shown). The bars describe the percentage of nearby opposite-spin sites when the central atom is spin-up (green) and spin-down (coral). (e)-(j) Lower panel: The $d$-orbitals PDOS of the manganese atom in the upper panel, respectively. (e) The NM model with a single atom in a unit cell induces a metallic gapless phase. (g)-(i) In the PM-SQS structure, the magnitude of moments in different local environments varies in a small range, and there is no accidental band gap closure for each individual atom. The inset shows the central atom’s environment corresponding to $n^{\uparrow}=5,3,2$ ($n^{\downarrow}=1,3,4$). (f) and (j) FM and AFM motifs.}
\end{figure*}

$\langle \Phi_{\alpha} \rangle = 0$ is quite helpful for making an approxiamtion. Although the magnetic moments are $360^\circ$ randomly aligned in real PM phases, we can still assume collinear spin configurations in calculations, as long as the parallel and antiparallel interactions are canceled when summing over clusters. The choice of the easy axis depends on the crystalline symmetries and may require testing. For example, van der Waals magnets usually exhibit large magnetic anisotropy energy (MAE), so their spin moment orientations can be initially set along the easy axis, and the amplitudes are free to evolve during the relaxation (flipping disallowed). But in certain scenarios ($e.g.$ Mn$_{3}$Sn has a noncollinear but coplanar order \cite{nakatsuji2015large}), we can adopt three (or more) in-plane directions as input to generate SQSs.

\begin{figure*}[t]
\includegraphics[width=0.95\linewidth]{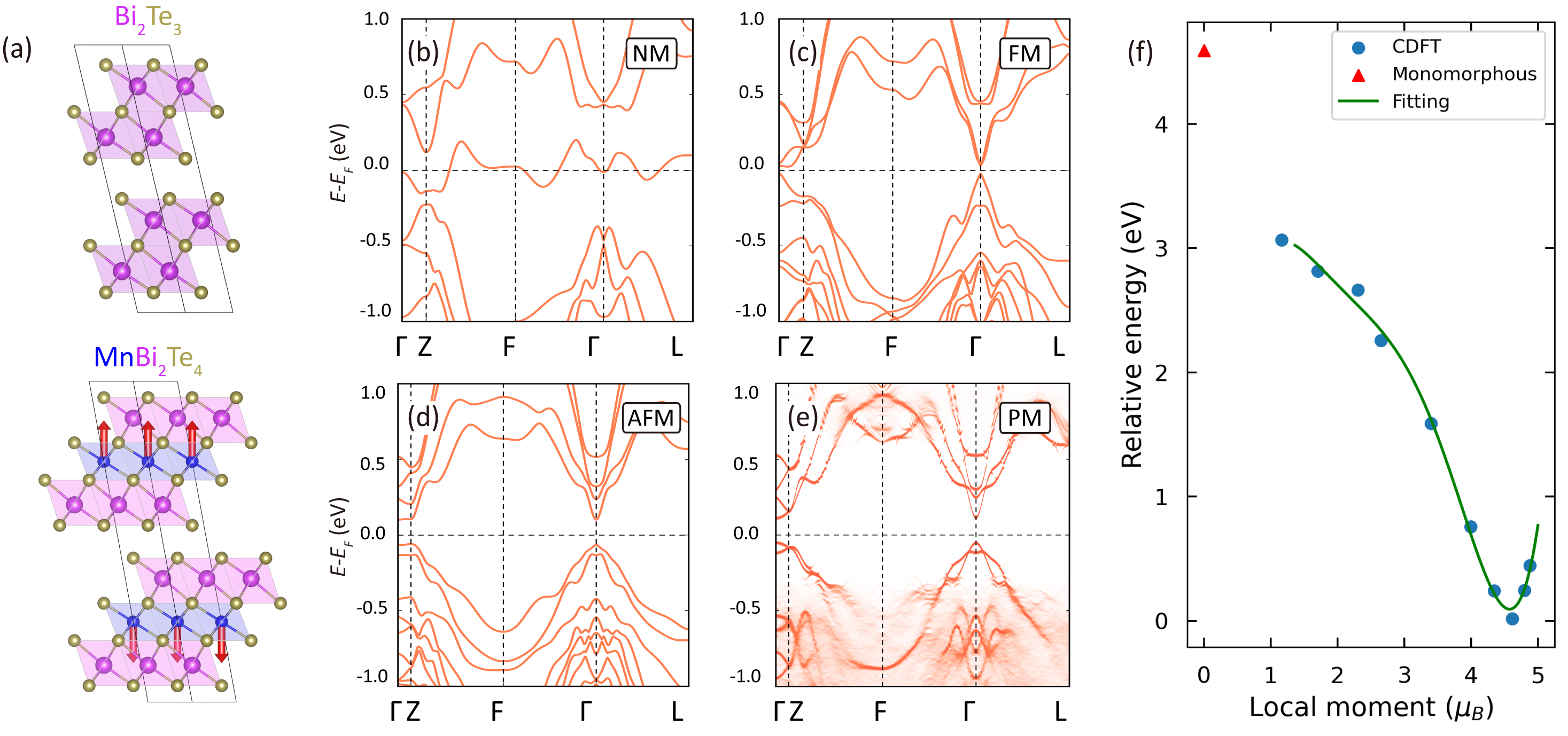}
\caption{\label{fig2}(a) Crystal structure of Bi$_{2}$Te$_{3}$ and MnBi$_{2}$Te$_{4}$ with an $A$-type AFM configuration. (b)-(d) Band dispersions of the (b) NM, (c) FM, and (d) AFM configurations. For the calculations of NM and FM phases, we use a single SL which causes double Brillioun zone length along $\Gamma-Z$. (e) The spectral functions of the PM phase by the polymorphous description with randomly distributed local moments. The spin-orbit coupling is included. (f) Total energy as a function of constrained local moments with an AFM configuration. A polynomial function is used to fit the total energy evolution.}
\end{figure*}

($iii$) \textit{Central limit theorem (CLT).} A basic intuitive understanding behind the PM phase is that there could exist multiple local environments rather than a single environment ($e.g.$ a cation can be surrounded by $n^{\uparrow}$ spin-up ions plus $n^{\downarrow}$ spin-down ions, but $n^{\uparrow, \downarrow}$ can differ at different sites). If we draw out a fixed-size supercell sample $i$ randomly from a real PM phase and count the number of cations that are surrounded by ($n^{\uparrow/ \downarrow} = 0,1,2,3,\dots$), we will have a distribution $F_{i}(n^{\uparrow/\downarrow})$, representing the probability of finding an ion with $n^{\uparrow/\downarrow}$. Repeating this \textit{independent and identically distributed} ($i.i.d.$) process, we can have
\begin{equation}
    \lim_{m\rightarrow\infty} \frac{1}{m} \sum_{i}^{m} F_{i}(n^{\uparrow/\downarrow}) \rightarrow \mathcal{N}(\mu,\sigma^{2}/\sqrt{m})
\end{equation}
where the right side $\mathcal{N}(\mu,\sigma^{2}/\sqrt{m})$ is a normal distribution with the mean value $\mu=0.5$. The finite variance of the real PM phase is set as $\sigma^{2}$. In density functional theory, it is inevitable to introduce spurious spatial correlations by periodic structures, but CLT hints at an underlying principle that, the supercell with a normal distribution of $F(n^{\uparrow/\downarrow})$ is optimal by technically minimizing the periodic errors. An example will be given in the next section (Fig. \ref{fig1}).

\section{\label{results}Results}
In this section, we will employ a comparative study with different descriptions on MnBi$_2$Te$_4$, which has received much attention recently \cite{10.1063/5.0059447,PhysRevLett.122.107202,YLChen,PhysRevX.11.011039,PhysRevB.103.L121112,PhysRevMaterials.5.024207}. This is the simplest example of a false-positive metal under the NM monomorphous method, enforced by an odd number of electrons within a unit cell and Kramers spin degeneracy.

As shown in Fig. \ref{fig2}(a), MnBi$_2$Te$_4$ crystalized in a van der Waals structure (space group R$\bar{3}$m) with a MnTe layer sandwiched by a quintuple layer (QL) Bi$_2$Te$_3$, forming a septuple layer (SL). At low temperatures, MnBi$_2$Te$_4$ exhibits an AFM ground state, where spin moments in each SL are FM coupled and point out of planes. With spin-orbit coupling effect induced band inversion, the system is a $\mathbb{Z}_2$ AFM TI with gapped (001) surface states. In the two-dimensional thin films, the system is predicted to show alternating behaviors between the QAH (odd number of SLs) and axion insulator states (even number of SLs) \cite{PhysRevLett.122.107202}. However, the (001) surface states, which should be gapped by magnetism, is experimentally observed to be gapless in the low-$T$ AFM phase. This unresolved issue hinders the observation of the AFM-PM phase transitions.

In the high-$T$ PM phase, MnBi$_2$Te$_4$ is expected to share many similarities as Bi$_{2}$Te$_{3}$. A naive but intuitive physical picture is as follows. Three-dimensional MnBi$_2$Te$_4$ will restore to a $\mathcal{T}$-preserved $\mathbb{Z}_2$ TI with the gapless (001) surface states. In the thin film, a low-$T$ QAH insulator is expected to be a QSH insulator or a trivial insulator in the high-$T$ limit \cite{PhysRevB.81.041307}.

Luckily, ARPES measurement found the temperature-dependent transition signal, albeit rather less salient. The (001) surface states kept gapless in both the AFM and PM phases, and the bulk gap was unchanged \cite{YLChen}. However, it was clearly shown that the two conduction bulk bands merged into one band at some specific $k$ path.

\subsection{Bulk}
The NM monomorphous approach, $i.e.$, the magnetic moment of Mn is set to zero within a primitive cell, is first considered to obtain the PM electronic structure of MnBi$_2$Te$_4$ (Fig. \ref{fig1}(a,e)). The unit cell consists of one SL that contains a single Mn atom. Guaranteed by space-time $\mathcal{PT}$ symmetry, each energy band is at least double-degenerate (Fig. \ref{fig2}(b)). Note that a single Mn atom has five 3$d$ electrons, so the total number of electrons within a primitive cell is odd. This indicates that the Fermi level must cut through at least one band, rendering a symmetry-enforced semimetal. Such a semimetal phase cannot be avoided by simply doubling the primitive cell, because it would fold each band at the boundary of the reduced Brillouin zone (BZ) rather than open a gap (Appendix Fig. \ref{fig4}(a)). Additionally, even if inversion symmetry is broken, it still exhibits a semimetal for Kramers degeneracy at time-reversal invariant wavevectors. Therefore, the oversimplified NM monomorphous model fails to predict the correct PM electronic structure of MnBi$_2$Te$_4$ due to the spurious high symmetry.

In the earlier DFT studies \cite{PhysRevB.87.195139}, the PM phase has also been naively treated as an AFM or FM phase (since the PM phase is composed of numerous microscopic AFM and FM configurations), which breaks not only the local $\mathcal{T}$ symmetry but also the global $\mathcal{T}$ symmetry. From the viewpoint of the distribution of spin configurations $F(n^{\uparrow/\downarrow})$ (the nearest neighbor shell), they could also be classified as monomorphous descriptions for identical local environments. The nearest neighbor sites of all cations have the maximum spin-similarity, $i.e.$, $F(n^{\uparrow/\downarrow} = 0) = 1$ that disobeys CLT. While the false-positive metallic prediction is avoided under the AFM/FM configuration (Fig. \ref{fig1} (f,j)), some characteristics of the band structure are significantly different from those in the real PM phase, as we discuss next.

In the polymorphous framework, the PM band structure can be accommodated by building a $4\times4\times2$ SQS supercell (224 atoms), where local moments are randomly distributed (inset of Fig. \ref{fig1}(c)). The distribution of spin configurations $F(n^{\uparrow/\downarrow})$ (the nearest neighbor shell) satisfies CLT. Strictly speaking, all the original space group symmetries cannot survive. We also check that any unnecessary microscopic translational symmetry imposed by the supercell is avoided. In Fig. \ref{fig1}(g-i), each individual Mn$^{2+}$ motif provides a finite band gap, rendering an overall bulk insulating phase. This is the most significant improvement compared to the NM monomorphous approach, consistent with ARPES measurements. 

Fig. \ref{fig2}(e) shows the DFT-calculated spectral functions in the SQS, which are unfolded into the BZ of a $1\times1\times2$ cell in order to directly compare with the AFM bandstructure (Fig. \ref{fig2}(d)). The fuzziness of the spectral density in the long-wave vector region reflects the degree of retention of crystalline symmetry. At a glance of the dispersion and gap, the energy band looks very similar to that of the AFM ground state. The energy gap size is nearly the same at $\Gamma$ ($\sim0.18 $ eV). Since ARPES did not observe any band gap closing and reopening, one may ask if this system is still topologically nontrivial. Although in this case any symmetry-based indicators cannot be defined, band inversion is a rough way to address this question. In Appendix Fig. \ref{fig5}, we show the valence and conduction band weights ($p$ orbitals of Bi and Te) at $\Gamma$ are also inverted in the PM phase. Thus, the PM phase of MnBi$_2$Te$_4$ belongs to a globally $\mathcal{T}$-preserved ``topological insulator'' \cite{YLChen}.

The evidence of the transition from the low-$T$ AFM to high-$T$ PM is reported to be two conduction bands merging at $Z$ (0,0,0.5) \cite{YLChen}. This merging effect leads to a four-fold degeneracy, which is captured by our polymorphous calculations. The reason behind it is physically insightful. The PM-SQS can be regarded as a composition of the amount of the AFM and FM $1\times1\times2$ unit cells. However, in the AFM and FM phases, no such nonsymmorphic symmetry exists to protect band crossings at $Z$ (Fig. \ref{fig2}(c,d)). In the mesoscopic scale of the PM phase, the translation, rotation, and $\mathcal{T}$ symmetries are recovered, but ARPES still detects the $1\times1\times2$ cell because of the presence of the local moments. Hence, this band merging phenomenon originates from the band folding effect at the BZ boundary when the system turns to the PM phase.

It is also known that the calculated energy difference reflects the stability of the phase or magnetic interactions \cite{PhysRevB.72.195210}. The energy of the PM-SQS is only 4.83 meV/Mn higher than that of the AFM ground state and 2.48 meV/Mn higher than that of the FM phase. To some extent, it corresponds to a low Néel temperature ($T_{N}$ $\approx$ 25 K). In sharp contrast, the total energy of the NM monomorphous phase (4.5 eV/Mn) is around 1000 times higher than that of the polymorphous one, which apparently deviates from physical reality. The inevitable omission of the Zeeman splitting of Mn atoms leads to such a huge difference, revealed by a CDFT calculation (Fig. \ref{fig2}(f)). When decreasing the magnitude of local moments under the AFM configuration, the total energy increases gradually. A jump occurs at $\vert \bm{s} \vert$ = 0 $\mu_{B}$ (NM case) because $\mathcal{PT}$ symmetry-enforced half-filling further lifts the band to Fermi level. To sum up, our results indicate the validity of the PM-SQS configurations when describing bulk states.

\subsection{Thin film}
\begin{figure*}[t]
\includegraphics[width=0.95\linewidth]{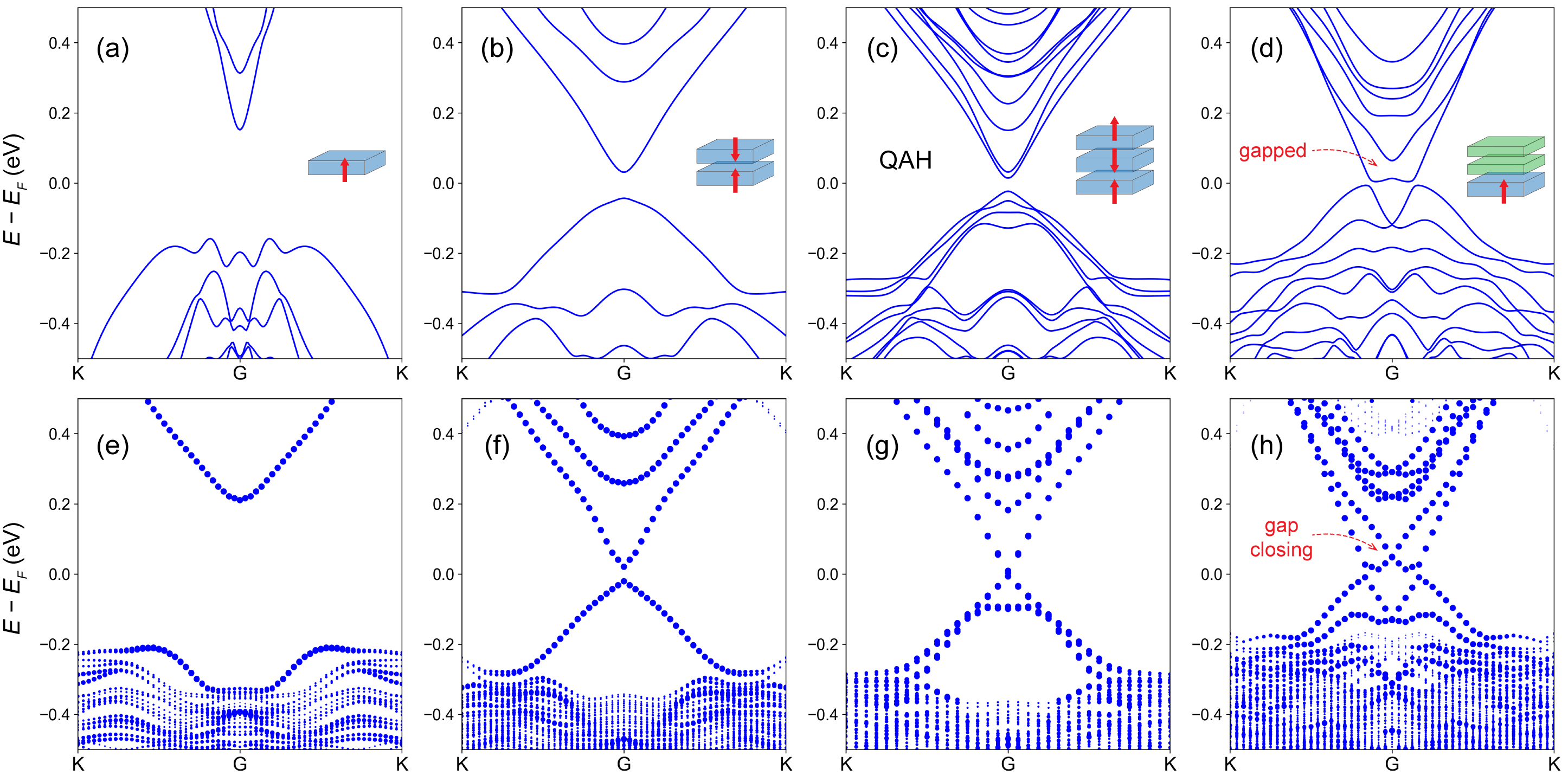}%
\caption{\label{fig3} (a-d) Calculated band structures of the thin film 1 SL, 2SL, 3SL, and SL-2QL, respectively. The AFM ground state is assumed. The 3SL is a QAH insulator.  (e-h) Band structures in the PM phase. Spectral weights are shown as the size of blue dots.}
\end{figure*}
Next we focus on the thin films of MnBi$_2$Te$_4$. For each structure considered in Fig. \ref{fig3}, we build a $4 \times 4$ SQS and unfold it back to the unit cell for simulating the PM phase. According to the Dirac equation, magnetization-induced $\mathcal{T}$-broken mass term will open a gap. However, the hybridization between top and bottom surfaces is also capable of gap opening. The polymorphous view is helpful in diagnosing the gap-opening mechanism.

On the other hand, when increasing the number of QLs/SLs, in Bi$_{2}$Te$_{3}$, there exists an oscillation behavior between the trivial insulator and the QSH insulator \cite{PhysRevB.81.041307, PhysRevLett.123.096401}; in MnBi$_{2}$Te$_{4}$, there exists an oscillation behavior between the axion insulator and the QAH insulator \cite{PhysRevLett.122.107202}. Although these terminologies are ill-defined in the PM phase, the polymorphous band structure can provide a basic picture of these properties.

From 1 SL to 3 SLs, the band structures of the AFM and PM phases are shown in Fig. \ref{fig3}(a-c, e-g). In 1 SL, the finite-size effect overwhelms the magnetization, so both phases show a large gap of around 0.4 eV. Upon increasing thickness up to 2 SLs, the AFM order is built with $\mathcal{PT}$ symmetry. With less hybridization, two-SL-MnBi$_{2}$Te$_{4}$ yields an axion insulator state with zero Hall plateau \cite{PhysRevLett.122.107202}. In comparison, the general energy spectra of the PM phase differ from those of the AFM phase with a much smaller energy gap of 32 meV. In 3 SLs, the AFM phase shows the QAH effect with a 74 meV gap, while the PM phase shows a gap of nearly 9 meV. Starting from 2 SLs, we could expect that the band gap of the PM phase keeps decreasing but without vanishing, analogous to its NM counterpart \cite{PhysRevB.81.041307}.

Note that MnBi$_{2}$Te$_{4}$ is a vdW compound, which enables a series of assembling structures with Bi$_{2}$Te$_{3}$. Thus, we also investigate an SL-QL-QL thin film. In the ground state (Fig. \ref{fig3}(d)), the previous studies have shown via the layer projection of the wave function that the energy gap at $\Gamma$ at $E-E_{F} = $ 0.02 eV, 0 eV, and $-$0.1 eV have different physical origins \cite{PhysRevB.102.035144, klimovskikh2020tunable, shikin2022electronic}. To be concrete, the gap at 0.02 eV [between the lowest conduction band and the second lowest conduction band] arises from the bottom SL and is associated with the magnetic exchange effect that breaks $\mathcal{T}$ symmetry and lifts the degeneracy; the gap at 0 eV is caused by the finite size effect. When the slab is thick enough, the hybridization between the top and bottom surface states is negligible, and the gap will vanish; the energy bands near $-0.1$ eV [the first two highest conduction bands] are dominant by the top NM QL, which approximately retains $\mathcal{T}$ symmetry because of the vanishingly small proximation effect. While the inversion symmetry is broken, this gapless Dirac point is still permitted but buried in the bulk valence bands. Hence, when the system turns to the PM phase [Fig. \ref{fig3}(h)], we find that only the gap at 0.02 eV closes while the latter two are essentially irrelevant to the magnetization, so they keep gapless/gapped. Our results reflect how the magnetic disorder restores the global $\mathcal{T}$ symmetry and then affects the gap.

\section{Discussion and Conclusion}
So far, we have analyzed the insulating system where each unit cell contains an odd number of electrons. One may ask if such a false-positive metallic prediction can be avoided in a system with an even number of electrons, so that the low-energy bands near the Fermi surface are guaranteed to be correct by the monomorphous description. It is accidentally true except in cases when crystalline symmetry plays a role. When the space group has a nonsymmorphic symmetry to accommodate the four-fold degeneracy, the system with $4n+2$ electrons is again inevitably in a metallic state under the monomorphous description.

In summary, we comprehensively investigated the PM electronic structure of bulk (3D) and thin films (2D) of topological insulators, exemplified by  MnBi$_{2}$Te$_{4}$. First, we found a band merging effect at the BZ boundary with negligible change in the bulk gap. Second, the surface gap opened by magnetization will vanish in the PM configuration while the hybridization gap does not change. It is worth noting that an open question regarding 1D edge states is left here, which may provide ingredients for the study of topological Anderson insulators \cite{PhysRevLett.102.136806}.

In the framework of single-particle mean-field DFT, performing the calculations with oversimplified approximation usually leads to unrealistic predictions, especially when the local disorder effect needs to be considered. In magnetic semiconductors where the local moments survive and distribute disorderly, a polymorphous description is required. Such an approach takes into account the otherwise neglected local symmetry breaking, as well as the correct exchange interaction. Our work provides a guiding principle to analyze the underlying physical properties of the PM phases, which will facilitate the investigation of the magnetic phase transition and topological magnets.

\begin{acknowledgments}
Y. Zhao thanks discussion with Shuolong Yang and Binghai Yan. This work was supported by the National Key R\&D Program of China (Grant No. 2020YFA0308900 and 2019YFA0704900), Guangdong Provincial Key Laboratory for Computational Science and Material Design (Grant No. 2019B030301001), Shenzhen Science and Technology Program (Grant No. RCJC20221008092722009), the Science, Technology and Innovation Commission of Shenzhen Municipality (Grant No. ZDSYS20190902092905285) and Center for Computational Science and Engineering of Southern University of Science and Technology.
\end{acknowledgments}

\appendix
\section{Convergence test of the supercell size}
\begin{figure*}[t]
    \centering
    \includegraphics[width=0.9\linewidth]{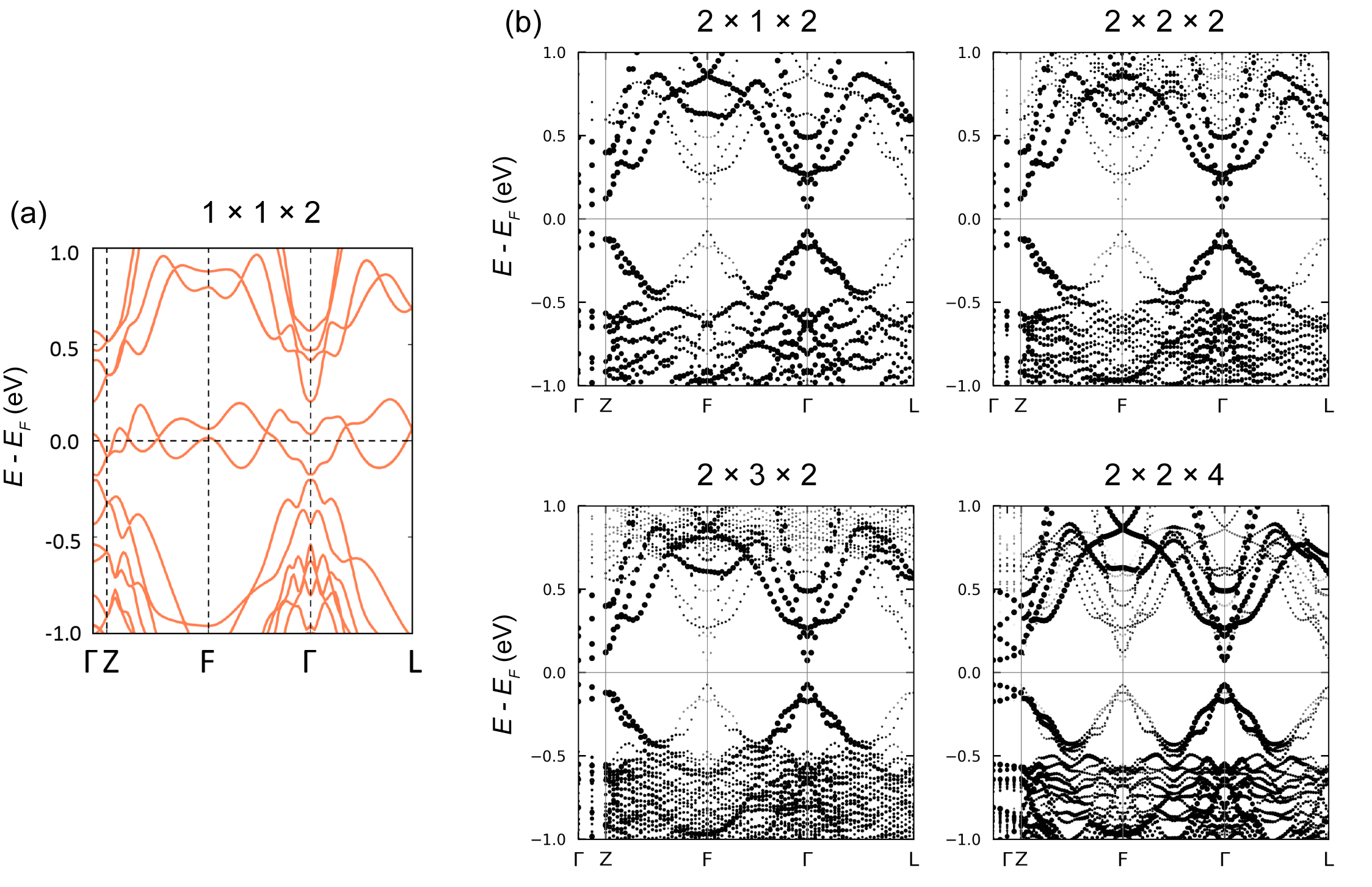}
    \caption{(a) Band structure under the monomorphous description. (b) Band structures under the polymorphous description by different size supercells, unfolded to a $1 \times 1 \times 2$ cell.}
    \label{fig4}
\end{figure*}

To rule out other possibilities of the band spectra, we also implement the calculations under different sizes of supercells with randomly distributed spin moments. There are three aspects to judge the convergence: band gap, dispersion, and degeneracy at high symmetry points. As shown in Fig. \ref{fig4}(a), the energy bands under the NM monomorphous description are double-degenerate at the Brillouin zone boundary ($Z$ and $L$). In principle, it can be used to judge the convergence of the supercell since two descriptions exhibit the same global symmetries. In Fig. \ref{fig4}(b), the band dispersion and unvanished gap obtained from other supercell sizes are close to the $4 \times 4 \times 2$ one (Fig. \ref{fig2}(e)). The band degeneracy at $Z$ point is quite robust, but the energy near $F$ and $L$ depends on the choice of the supercell.

To be concrete, while the $2 \times 1 \times 2$ cell accidentally achieves the degeneracy at $L$, the $2 \times 2 \times 2$ and $2 \times 2 \times 4$ cells fail to show that. Besides, the bands near $F$ are evident to show the importance of the supercell size on the $ab$ plane. Among them, only the $2 \times 3 \times 2$ and $4 \times 4 \times 2$ supercells are satisfactory for the dispersion near the $F$ point.

\begin{figure}[t]
\includegraphics[width=1\linewidth]{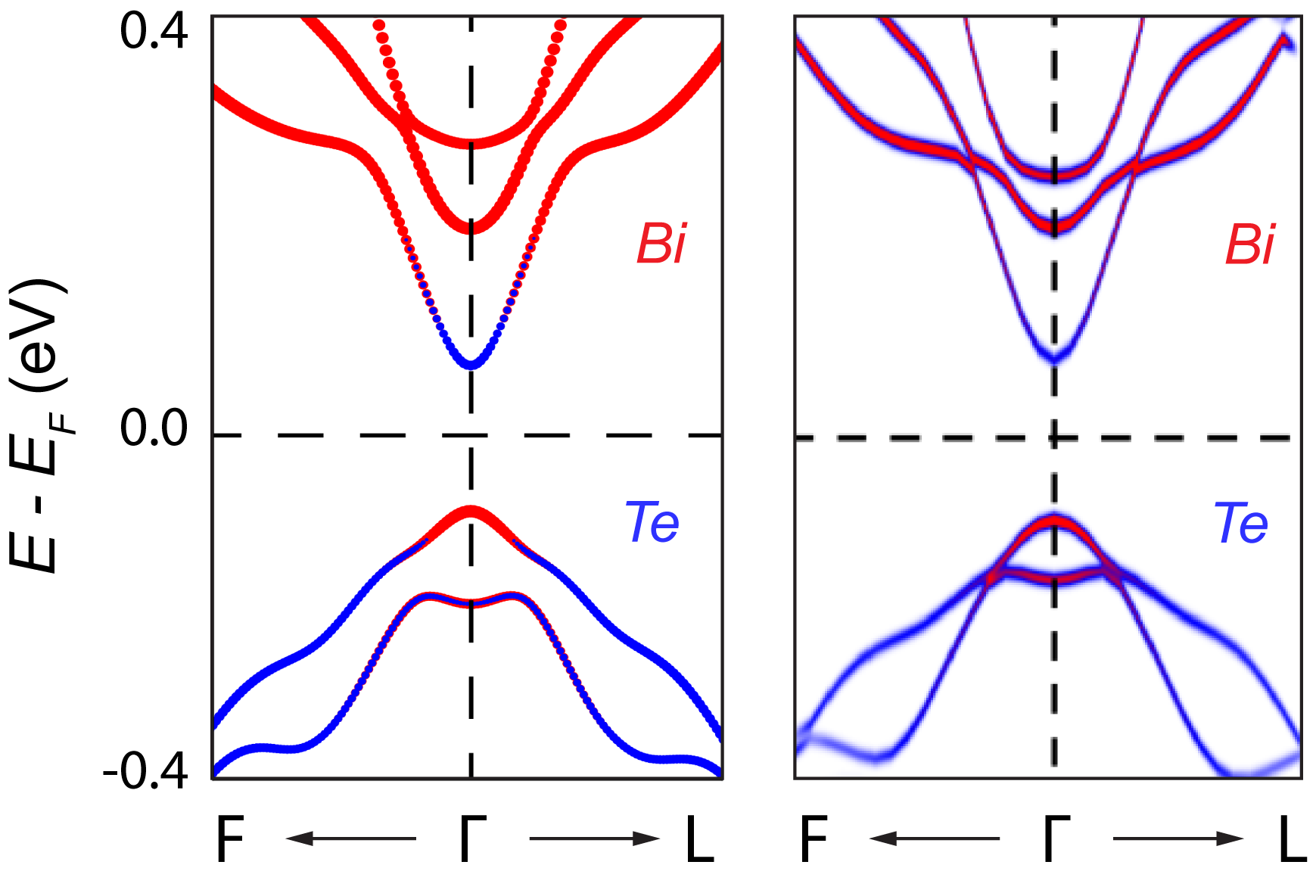}%
\caption{Projected energy spectrum (Bi-$p$ and Te-$p$ orbitals) in the AFM phase (left) and PM phase (right). An inverted band order observed at $\Gamma$ indicates a ``PM topological insulator''.}
\label{fig5}
\end{figure}

%

\end{document}